\input{aipcheck.tex}

\documentclass[final,numberedheadings]{aipproc}

\gdef\apjl{ApJ}
\gdef\aj{AJ}
\gdef\physrep{Phys.\ Rep.}
\gdef\apj{ApJ}
\gdef\mnras{MNRAS}
\gdef\aap{A\&{}A}
\gdef\apjs{ApJS}
\def\etal{{\it et~al.}}

\layoutstyle{6x9}

\begin{document}

\title{The Warm Spitzer Mission: Prospects for Studies of the Distant Universe}

\keywords{Spitzer Space Telescope, infrared astronomical observations, external galaxies, 
quasars, distances, redshifts}
\classification{95.85.Hp, 98.62.Py, 98.80.Bp, 98.54.Aj}

\author{Pieter van Dokkum}{address={Department of Astronomy,
Yale University, New Haven, CT 06520-8101, USA},email={pieter.vandokkum@yale.edu},}

\iftrue
\author{Asantha Cooray}{
address={Center for Cosmology, Dept of Physics and Astronomy, University of
California, Irvine, CA 92697, USA},email={acooray@uci.edu},}

\author{Ivo Labb\'e}{
address={Carnegie Observatories, 813 Santa Barbara St, Pasadena, CA 91101, USA},
email={ivo@ociw.edu},}

\author{Casey Papovich}{
address={Steward Observatory, University of Arizona, 933 North Cherry Ave,
Tucson, AZ 85721, USA},email={papovich@as.arizona.edu},}

\author{Daniel Stern}{
address={Jet Propulsion Laboratory, California Institute of Technology,
Pasadena, CA 91109, USA},email={stern.thisvi.jpl.nasa.gov}}

\fi

\begin{abstract}

IRAC excels at detecting distant objects.
Due to a combination of the
shapes of the spectral energy
distributions of galaxies and the low background achieved from space,
IRAC reaches greater depth in comparable exposure time
at 3.6 and 4.5 $\mu$m
than any ground- or space-based facility currently can at 2.2 $\mu$m.
Furthermore, the longer wavelengths probed by IRAC enable
studies of the rest-frame optical and near-infrared light of
galaxies and AGN to much higher redshift than is possible from the
ground. This white paper 
explores the merits of different survey strategies for
studying the distant universe during the warm mission.
A three-tiered approach serves a wide range of science goals
and uses the spacecraft effectively:
1) an ultra-deep survey of $\approx 0.04$ square degrees to
a depth of 250\,hrs (in conjunction with an HST/WFC3 program),
to study the Universe at $7<z<14$;
2) a survey of  $\approx 2$ square degrees
to the GOODS depth of 20 hrs, to identify luminous
galaxies at $z>6$ and
characterize the relation between
the build-up of dark matter halos and their constituent
galaxies at $2<z<6$, and 3) a 500 square degree survey to the
SWIRE depth of 120\,s, to systematically study large scale
structure at $1<z<2$ and characterize high redshift
AGN. One or more of these
programs could conceivably
be implemented by the SSC, following the example of the Hubble
Deep Field campaigns.
As priorities in this field continuously shift it is also
crucial that
a fraction of the exposure time remains unassigned,
thus enabling science that will reflect the frontiers
of 2010 and beyond rather than those of 2007.

\end{abstract}

\date{\today}
\maketitle

\section{Introduction}

Infrared observations are crucial for the study of distant galaxies.
While blue star forming galaxies can be routinely identified
to $z\sim 6$ and beyond using
optical selection techniques and follow-up spectroscopy
(e.g., \citet{Steidel1996a, Steidel1996b, Steidel2004, Kodaira2003, Ouchi2005,
Stark2007a, Dow-Hygelund2007}), measuring their
masses and star formation histories requires access to their
rest-frame optical light (see, e.g., \citet{Shapley2001,
Papovich2001}). Furthermore, it has become clear
that optical samples  miss a substantial fraction
of the high redshift galaxy population.
Near-infrared surveys
have discovered substantial numbers of UV-faint red galaxies
(\citet{Daddi2000, McCarthy2001, Labbe2003, Franx2003})
and it appears that these objects dominate the $z=2-3$ cosmic stellar mass
density at the high-mass end (\citet{vanDokkum2006,
Marchesini2007}). In addition,
surveys at mid-infrared, sub-mm, and radio wavelengths have found
highly obscured galaxies, which emit the bulk of their luminosity
at IR wavelengths (e.g., \citet{Barger2000, Blain2002}) and may contribute
substantially to the global cosmic star formation rate.

The infrared capabilities of the
Spitzer Space Telescope have greatly enhanced our understanding of
the high redshift Universe. MIPS and IRS are rapidly advancing our knowledge
of IR luminous galaxies, such as obscured
Active Galactic Nuclei (AGN) and star burst
galaxies harboring large amounts of dust (see, e.g.,
\citet{Marleau2004, Dole2004, Yan2005a, Houck2005, LeFloch2005,
Frayer2006, Papovich2006}). However, MIPS is not able to study
``normal'' galaxies out to very high redshift: at redshifts as
low as $z\sim 3$ a galaxy has to have a star formation rate exceeding
$\sim 200\,M_{\odot}$\,yr$^{-1}$
to be detectable at 24 $\mu$m even in the deepest (10 hr)
images, and many times higher to be detected
at higher redshift or longer wavelengths.

\begin{figure}[t]
\resizebox{1.0\textwidth}{!}{\includegraphics{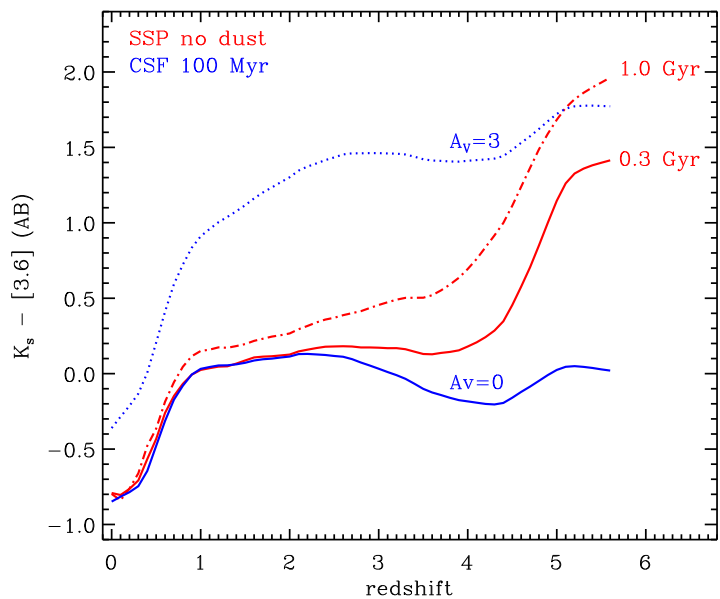}}
\caption{
Expected $K-[3.6]$ color of galaxies versus redshift from \citet{Bruzual2003} 
stellar population synthesis models. The bluest
dust-free galaxies have observed $K-[3.6]$ colors $>0$ at
most redshifts $>1$ (blue solid line). Dusty galaxies at
$z>1$ (blue dotted line) and galaxies with old stellar populations at
$z>4.5$ (red solid and dash-dotted line) have much redder colors,
reaching $K-[3.6] \sim 2$, which implies that they are {\it much} easier to
detect with IRAC than with ground-based near-infrared cameras.
}
\label{K36.plot}
\end{figure}

\begin{figure}[h]
\resizebox{0.95\textwidth}{!}{\includegraphics{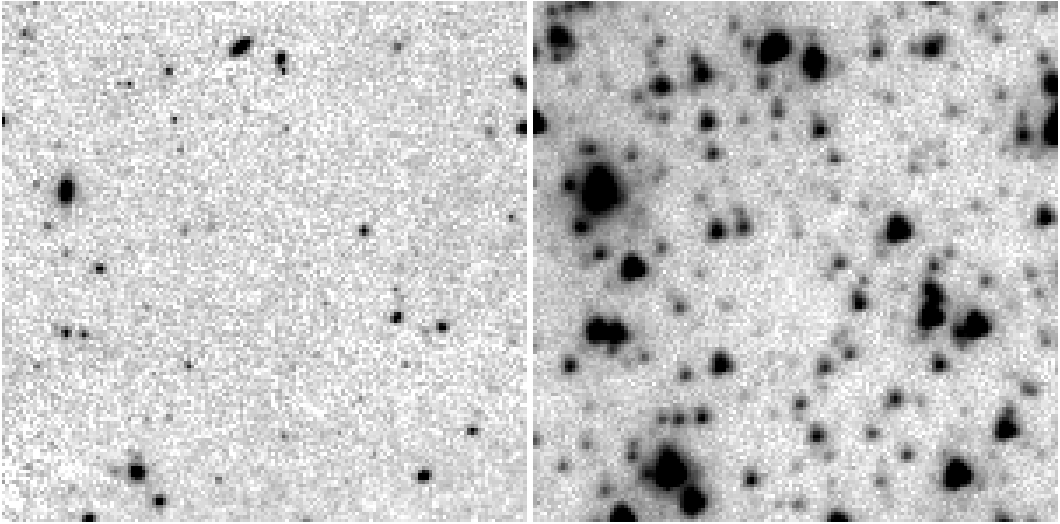}}
\caption{Comparison of ground-based $K$ (left) with
Spitzer 3.6\,$\mu$m (right), for a $1.5' \times
1.5'$ patch in the GOODS CDF-South field.
The $K$ band data were taken with ISAAC on the VLT and
are of very high quality. Per-pixel exposure times were $7$ hrs in $K$
and 20 hrs in the 3.6\,$\mu$m band. Despite a very large investment
of VLT time in this field (double that of IRAC due to the smaller field
of view, for a total of 288 hrs as of writing) the $JHK$ depths in GOODS 
South are poorly matched to the IRAC depth.}
\label{K_ch1.plot}
\end{figure}

IRAC, by contrast, excels at detecting distant galaxies of any kind,
due to a combination of the shape of their spectral energy distributions (SEDs) and the
low background achieved from space. As illustrated in Fig.\ \ref{K36.plot}
the bluest galaxies at $z>1$ have a $K-3.6$ color of $\sim 0$ in AB units. As
IRAC can reach the same AB depth as a ground-based 4m telescope
about 20 times faster, Fig.\ \ref{K36.plot} implies that any $z>1$ object detected with
a 4m telescope in the $K$ band can be detected with IRAC in 5\,\% of the exposure time.
For intrinsically red sources this difference is,
of course, even larger: dusty galaxies at any redshift and old
galaxies beyond $z\sim 4.5$ typically have $K-[4.5] \sim 2$, and
for these objects IRAC is a factor of 800 faster than a ground-based
4m telescope!
The  difference in depth achievable from the ground and
from space is illustrated in Fig.\ \ref{K_ch1.plot},
which compares a region of the CDF-South
field in $K$ to the
corresponding 3.6\,$\mu$m image. The (VLT) near-IR data in CDF-South
are among the best available anywhere in the sky,
and yet they are obviously
not well matched in depth to the IRAC data.
\vspace{0.2cm}

For studies of the distant Universe, the key advance allowed by IRAC is
not simply survey speed, but the abilitiy
to study the rest-frame optical and near-infrared light of
galaxies and AGN to much higher redshift than is possible from the ground.
As an example, at $z=7$
the $K$ band samples the rest-frame UV light of galaxies, which is
dominated by short-lived O and B stars, whereas the IRAC 4.5\,$\mu$m
band samples the rest-frame  $V$ band,
which provides information on Solar type stars and constrains
the age and mass of the bulk of the stellar population.

Major achievements with IRAC include:
measurements of the abundance of obscured QSOs
(\citet{Lacy2004, Treister2004, Stern2005, Cool2006});
identification of galaxy clusters and groups in the
redshift range $1<z<2$ (\citet{Brodwin2006}); identification
of massive galaxies with very low star formation rates at $z=2-3$
(\citet{Yan2004, Labbe2005});
determination of stellar ages and masses
of galaxies out to $z\sim 6$ (\citet{Eyles2005, Yan2005b,
Stark2007a}); confirmation and characterization of
galaxies at $z\sim 7.5$ (\citet{Egami2005, Labbe2006}); and possibly the
detection of fluctuations induced by first-light galaxies
containing a large fraction of population III stars
(\citet{Kashlinsky2005}).

Nearly all these results were driven by the short wavelength channels of
IRAC, as they are the most sensitive. In the warm mission, it will be
possible to extend these initial studies to wider areas and larger samples, 
as well as to fainter luminosities and higher redshifts.
Furthermore, very large programs
will enable entirely new science, in particular when combined with
planned extensive public near-infrared imaging surveys in the next
five years.

Here we describe a three-tiered survey program which could be conducted
over the course of the warm mission. The surveys
comprise ultra-deep observations in
a relatively small area,
a deep (20\,hr per pixel) program over a $2$
square degree area,
and a shallow (120\,s per pixel) program over a
$500$ square degree area.
These programs serve
as examples of science that can be done during the
warm mission; some other options are
briefly discussed in a separate Section.

\section{An Ultra-Deep Survey}
\label{ultradeep.sec}

\begin{figure}[h]
\resizebox{0.95\textwidth}{!}{\includegraphics{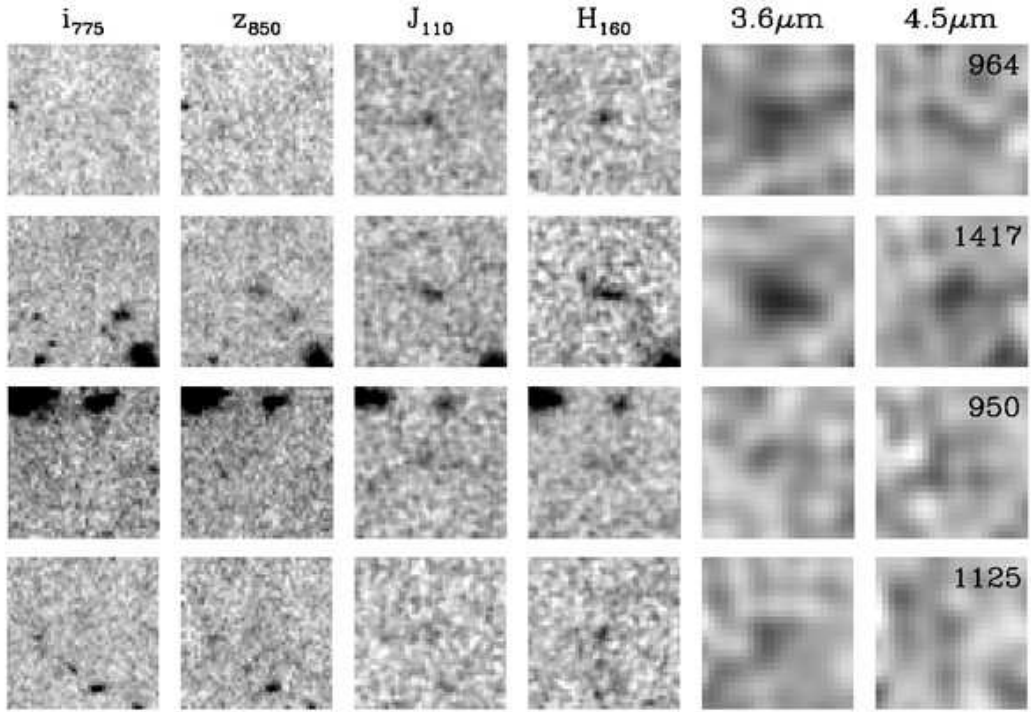}}
\caption{IRAC imaging of
$z$-dropouts in the Hubble Ultra Deep
Field, from \citet{Labbe2006}. In $\approx 46$ hrs two of these
faint z-dropouts are detected with IRAC, and two are marginally
detected. To characterize the $z\sim 7+$ galaxy population with IRAC
longer integration times and surveys over larger fields are needed.
}
\label{zdrop.plot}
\end{figure}

At redshifts above $z\sim 5$ the Balmer break shifts beyond the
observed $K$ band, and IRAC is the only instrument until JWST which
can provide reliable ages and masses of very high redshift galaxies.
As an illustration of the power of IRAC at very high
redshifts, Fig.\ \ref{zdrop.plot} shows
3.6 and 4.5 $\mu$m imaging of $z=7-8$ objects identified
in the Hubble Ultra Deep Field.
The integration time of $\approx 46$ hrs
per pixel was sufficient to robustly detect two of the four objects,
providing first estimates of the masses,
stellar ages, star formation rates, and dust content of these early
objects (\citet{Labbe2006}). Similarly, \citet{Egami2005}
used IRAC to constrain the stellar population of a lensed $z\sim 7$
galaxy.

The galaxy population at $z\sim 7$ may be responsible for
reionizing the universe and is of vital importance for
understanding feedback and metal production in the earliest stages
of galaxy formation. Much deeper IRAC
observations over a much wider area than the Hubble Ultra
Deep Field are needed to systematically survey the Universe at this
important juncture in its history.
Extrapolating from the \citet{Bouwens2006, BouwensIllingworth2006,
Labbe2006}) results, and taking the reduced area due to source confusion
into account (see below),
a survey over $\sim 150$\,arcmin$^2$ with a per-pixel
integration time of $\sim 250$\,hrs is needed to obtain a sample
of $\sim 100$ galaxies at $6.5<z<7.5$.with $>5\sigma$ IRAC
photometry. The total time required for this survey is $\sim
2500$\,hrs.\footnote{A $3\times 3$ pointing mosaic;
total survey times in this document include overheads.}

An ultra-deep survey would also offer the exciting prospect of
a first exploration of the $z\sim 10$ Universe, well in advance of JWST.
The depth achieved  is
$\sim 0.04\,\mu$Jy at 3.6\,$\mu$m ($\sim 27.4$ AB),
or $M_B \sim -22.6$ at $z=10$. The expected number of $z\sim 10$
objects is obviously very uncertain, but based on results to
$z\sim 7$ one may conservatively
expect to detect a handful of galaxies
at $9<z<11$ ($J$-dropouts), and $\sim 1$ object at $12<z<14$
($H$-dropouts, selected on the basis of their blue $[3.6] -[4.5]$
color and non-detection in HST/WFC3 $H$).\footnote{These
numbers are somewhat conservative as they
assume very blue SEDs redward of $\lambda_{\rm rest}
\sim 1400$\,\AA.} IRAC photometry of galaxies in this
redshift range provides very strong constraints on the
formation of the first stars.
If $z\sim 10$ galaxies experienced their first
star formation at this redshift, their $K$ through
4.5\,$\mu$m SEDs would
be powerlaws (with the powerlaw index an indication of dust
and metal content); if, on the other hand, these objects show
a pronounced break between the 3.6 and 4.5\,$\mu$m band their
spectra have a significant contribution of A stars and star
formation must have started several 100 Myr earlier,
at $z\sim 20$.

An ultra-deep campaign also offers the possibility of placing limits
on the frequency and nature of pair-creation SNe. These SNe are
thought to be the end states of very massive (150 -- 200\,$M_{\odot}$),
metal poor stars which may have existed in the early Universe
(\citet{Abel2002, Bromm2002}). The
peak brightnesses of such SNe are very uncertain, and could range
from $0.01 - 1\,\mu$Jy at $z=10$ (Scannapieco 2006). The rates are
also very uncertain, with estimates ranging from 1 -- 100
deg$^{-2}$\,yr$^{-1}$. With an optimized observing cadence 
the proposed survey probes the low peak brightness, high rate regime,
whereas a wide, shallower survey
probes the high peak brightness, low rate regime.

As illustrated in Fig.\ \ref{zdrop.plot} it is crucial to have supporting
near-infrared data that is well matched to the Spitzer depth -- in fact,
the IRAC data in isolation have very limited value. 
The near-IR data are needed to identify the high redshift
galaxies (by pinpointing their redshifted Lyman break at
1216\,\AA) and to obtain accurate photometry in the IRAC
images (by iteratively modeling the source distribution).
Obtaining sufficiently deep $K$ band data is extremely difficult,
even if one focuses on the bluest galaxies only.
Using the ``factor of 20'' rule of thumb a 250 hr IRAC depth
implies a per-pixel integration time of 5000 hrs on a 4m, or
1000 hrs on a 8-10m class telescope. Fortunately it will be
possible to reach the required depth
in $J$ and $H$ with HST/WFC3. Based on existing NICMOS data
in the Hubble Ultra Deep Field and the expected sensitivity
of WFC3, $\sim 9$ orbits are needed to match the
depth of a 250 hr IRAC observation. To cover an area
of 150 arcmin$^2$ in $J$ and $H$ would
require about 700 orbits. Given the importance of
supporting near-IR data an ultra-deep IRAC program
should probably only be undertaken in coordination
with an investment of HST time of this order.

A drawback to an ultra-deep field is the limited efficiency
of IRAC at faint flux levels due to crowding. At the GOODS depth
only $\sim 30$\,\% of pixels are uncontaminated
background, that is,
not affected by the wings of the PSFs
of identified sources. Source confusion is
not a hard limit, and can be greatly reduced with the use of
a prior image with better resolution (typically a $K$-band
image). However, confusion reduces the efficiency of IRAC
observations in two ways: the fraction of
the field in which good photometry can be done steadily diminishes
when going deeper, and the S/N increases slower than $\sqrt{t}$
due to the steadily rising ``background'' of PSF wings.\footnote{For
example, data in the Hubble Ultra Deep Field ($\approx 46$ hrs)
suggests that the depth increase compared to 1 hr is only 1.7
mag instead of the 2.1 mag expected from $\sqrt{t}$, even
after reducing the source confusion using available NICMOS near-IR
data (\citet{Labbe2006}).} At the time of writing, no results are yet
available from the deepest -- 100 hr per pixel -- region that has
been obtained with IRAC
so far (GOODS HDF-N); when these results are in it will be easier to
quantify the effects of crowding with integration times $> 50$
hrs per pixel.

Another drawback is that this type of science
in particular can be done with much
greater efficiency with JWST. There is little doubt that JWST
will image GOODS-sized fields (and larger), and that the depth
of IRAC data can be surpassed very rapidly indeed: quite apart
from its vastly superior PSF ($0.1'' - 0.2''$ at $3-5$\,$\mu$m)
the required exposure time
to reach a given point-source depth is about
three orders of magnitude shorter. Assuming a typical high
redshift galaxy size of $0.5''$ ($1.0''$)
FWHM and factoring in
the respective detector sizes, JWST/NIRCam can
cover large areas about 200 (40) times faster than Spitzer/IRAC.
Although this may limit the legacy value of an ultra-deep Spitzer survey,
such considerations have to be weighed against
the long lead time for JWST and the uncertainties associated
with any space mission.

\section{A Deep Survey Over 2 deg$^2$}
\label{deep.sec}

Although much larger than the original Hubble Deep Fields, the
$10' \times 15'$ GOODS
fields (\citet{Dickinson2004})
are too small to provide a fully representative sample
of the distant Universe: the correlation length $r_0$ of
massive galaxies is $\sim 8 h_{100}^{-1}$\,Mpc
(roughly independent of redshift), which is $\sim 8'$ at $z=2$
(e.g., \citet{Daddi2000, Somerville2004, Adelberger2005, Quadri2007}).
The GOODS fields are also too small for clustering studies
(except for populations with small $r_0$), and for studies
of the relation between galaxy properties and density.
The importance of sampling large volumes at high redshift
is dramatically illustrated by the identification
of structures of several tens of Mpc
up to $z\sim 6$ (e.g., \citet{Ouchi2005}).

Furthermore, the relatively small size of GOODS does not sample
the bright end of the luminosity function well, which means that
the brightest galaxies at high redshift are
missed even if the depth is sufficient to detect them.
As an example, the \citet{Bouwens2006}
$z=6$  luminosity function implies that only $\approx 5$
$L>3L_*$ galaxies at $5.5<z<6.5$
are expected in a $150$\,arcmin$^2$ area.
Although these bright examples may not contribute greatly to
the total luminosity density at these early epochs (see, e.g.,
\citet{Bouwens2006}), they may be accessible for morphological
studies with WFC3 and spectroscopic
follow-up with 20m -- 30m telescopes and JWST. 

Motivated by similar concerns, several programs are underway to
extend the area covered by deep ground- and space-based observations.
Examples are the $30' \times 30'$ Extended CDF-South (E-CDFS, aka
the GEMS field); the $50' \times 50'$ UKIDSS Ultra Deep Survey
(aka the Subaru/XMM deep field); the $10' \times 60'$ Extended
Groth Strip; and the $1.4^{\circ} \times 1.4^{\circ}$ COSMOS
field. All these fields have excellent supporting data, although
different fields have different strengths.
Current IRAC coverage of these fields varies. The E-CDFS and
the Groth Strip have both been covered with IRAC to
$\sim 3$ hr depth. The UDS will be done with IRAC to
$\sim 0.7$ hr depth in Cycle 4, and the
COSMOS field has relatively shallow ($\sim 0.3$ hr) IRAC
coverage over the entire field.

Given the large investments of ground- and space-based observatories
in these fields it seems likely that they
will continue to play important roles in
studies of the distant Universe.
New instrumentation
on existing telescopes (e.g., multi-object
near-IR spectrographs on 10m class telescopes and WFC3 on
HST) will likely be utilized in these fields, as well as future
telescopes (Herschel, ALMA, 20-30m telescopes, JWST).
There is therefore a strong legacy argument to be made
for covering several or all of these fields with substantially
deeper 3.6 and 4.5 $\mu$m imaging than is currently available.

The availability of near-IR imaging that is well
matched to the IRAC depth is crucial for correctly measuring the
IRAC fluxes and for determining photometric redshifts. Interestingly,
{\em none} of the fields mentioned currently has near-IR coverage
approaching the depth achieved in a few hours (per pixel) with IRAC.
However, this situation will change in the near future thanks to
ambitious public surveys with new
large field near-IR imagers on 4m class telescopes.
WFCAM on UKIRT  will cover the
Subaru/XMM deep field to a $5\sigma$ AB depth of $K=25$
(with additional $J$ and $H$) in the context of the UKIDSS
Ultra Deep Survey (\citet{Dye2006}).
UltraVISTA (an approved public survey on the soon to be commissioned
VISTA telescope)
aims to cover 1/3 of the COSMOS field to a depth of
$K=24.5$ and 1/3 to a depth of $K=25.6$ (with additional
$Y$, $J$, and $H$).
An IRAC depth of 20 hrs per pixel is well matched to the $K$ band depths
of UKIDSS/UDS and UltraVISTA, in the
sense that every $K$-detected source will have a
3.6\,$\mu$m $>5\sigma$ counterpart.

Covering the other two fields should also be a high priority. Their
areas are small compared to the UKIDSS/UDS and COSMOS
UltraVISTA fields --- which implies that the investment with
Spitzer would be relatively modest --- and they offer
qualitatively different legacy value. Covering
only the 0.7 deg$^2$ UDS field and the 0.8 deg$^2$
COSMOS/UltraVISTA field
to the GOODS depth would cost $\sim 6,000$ hrs, whereas
covering all four fields would require $\sim 7,500$
hrs.\footnote{In practice, it may be beneficial to vary the exposure
time within a field or between fields somewhat.}

The area and depth of such a $\approx 2$ deg$^2$ survey
should be sufficient to detect 1000s of galaxies at $5<z<8$.
At these redshifts IRAC uniquely samples the rest-frame optical 
emission beyond the Balmer break,
allowing measurements of star formation histories and stellar
masses (see, e.g., \citet{Labbe2006, Stark2007a}).
The intrinsic brightness of these objects implies that they
can be observed spectroscopically, either with existing telescopes
or with future 20m-30m telescopes and/or JWST. In combination
with the ultra-deep survey discussed above,
which samples the luminosity function at $L< L_*$,
the evolution of the
rest-frame optical luminosity function and the stellar mass
function can be accurately measured at $5<z<8$.

Furthermore, the survey will characterize
the relation between galaxies and the emerging large scale
structure over the redshift range $2<z<6$.
GOODS-depth IRAC observations over 2\,deg$^2$ would
allow characterization of the stellar populations
of several tens of thousands of red and blue galaxies
in this redshift range to low
stellar mass limits (e.g., $\sim 10^{10} M_{\odot}$ at $z=3$)
and accurately determine their density and evolution in
relation to their environment. The combination of clustering
and stellar population measurements is an extremely powerful
tool to determine the properties and
evolution of galaxies as a function
of halo mass (e.g., \citet{Adelberger2005,
Lee2006, Quadri2007}),
thus linking the hierarchical build-up of
dark matter halos to the formation and evolution of their
constituent galaxies.

Deep IRAC observations over such a large area
also offer intriguing possibilities for studies of
faint galaxies below the detection threshold.
If first-light galaxies during
reionization were to contain a substantial fraction of massive
population III stars then their redshifted rest-frame UV emission will be
present at IR wavelengths. While none of these sources will be individually
detected even in the ultra-deep survey described earlier,
the unresolved emission
will be clustered (as these sources are expected to trace the large-scale
structure at $z > 7$) and this clustering component
can be extracted to the extent
that any correlated systematics and noise sources are understood.
A first detection of such a clustered component in the unresolved
IRAC pixels in the first-look survey was interpreted as evidence for
massive population III stars (\citet{Kashlinsky2005}), although
this result is somewhat controversial (\citet{Cooray2007}).
A deep survey over 2\,deg$^2$ will make it possible to accurately
measure the clustering strength of the undetected sources, allowing
a direct comparison to model predictions for the clustering of
first-light objects.

\section{A Shallow Survey Over 500 deg$^2$}
\label{wide.sec}
 
Areas of several square degrees are sufficient to obtain representative
samples of the Universe at $z\leq 1$, but they are not large enough
for studies of extreme objects such as luminous quasars or high-redshift
galaxy clusters. Although the instantaneous field-of-view of IRAC is
small, it can do very efficient mapping over large areas of sky; as
an example, the SWIRE survey covered 49 deg$^2$ to a depth of 120\,s.
An order of magnitude larger survey than SWIRE would take $\sim 4000$
hrs and serve a wide range of science goals.
 
High redshift quasars can be efficiently identified by their relatively
flat mid-IR SEDs and their extremely red optical -- mid-IR colors (\citet{Cool2006,
Stern2007}). The clustering strength of these objects
will constrain the masses of their dark matter halos, and spectroscopic
follow-up will provide information on the build-up of supermassive black
holes and the interplay of star formation and nuclear activity in the
earliest phases of galaxy formation.  Quasars also provide useful probes
of the intervening universe;  indeed, the most distant quasars provide
some of our most powerful tools for probing the epoch of reionization
(\citet{Becker2001}).
 
\begin{figure}[h]
\resizebox{0.99\textwidth}{!}{\includegraphics{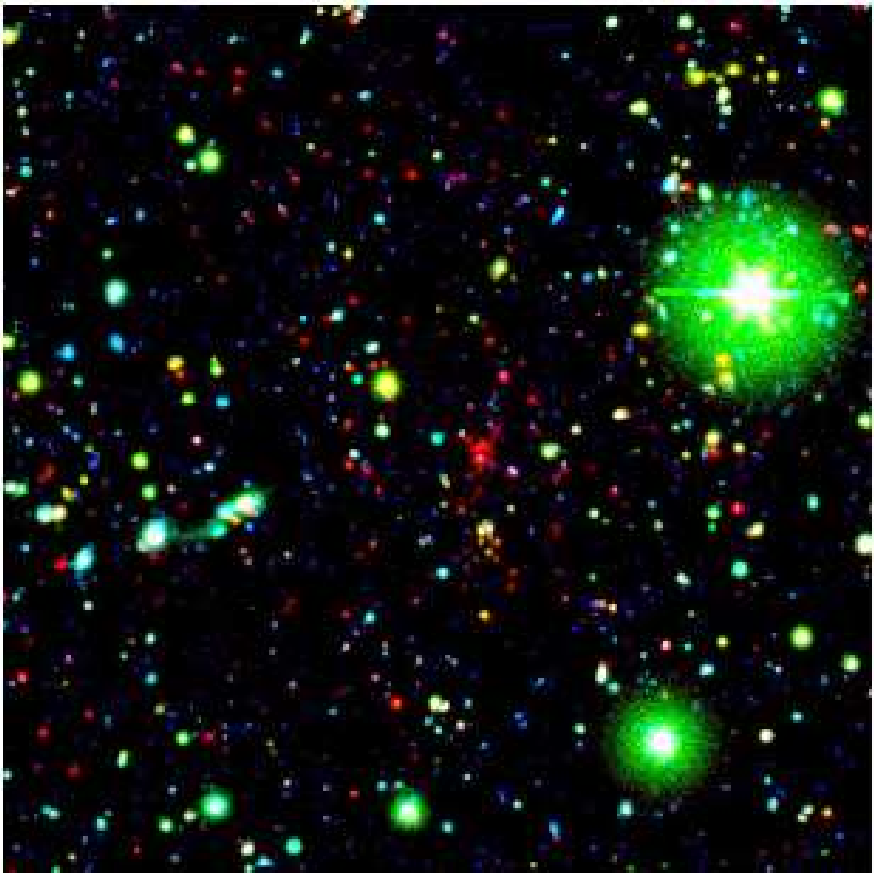}}
\caption{
Color composite of $B$, $I$, and IRAC 4.6\,$\mu$m images of a
galaxy cluster at $z=1.41$, from \citet{Stanford2005}.
The ground-based $B$ and $I$ images required several hours of
exposure time on a 4m telescope, but the integration
time for the IRAC 4.5\,$\mu$m image was only 90 seconds!
}
\label{clus.plot}
\end{figure}

Galaxy clusters can easily be identified out to $z\sim 2$ with IRAC in
integration times as short as a few minutes (see \citet{Eisenhardt2006}
and Fig.\ \ref{clus.plot}).
Based on the WMAP3 cosmology, a 500 deg$^2$ survey to the
SWIRE depth would provide $\sim 1500$ clusters at $1<z<2$ with masses
$>10^{14}\,M_{\odot}$, and a handful of extremely massive objects with
masses $>5 \times 10^{14}\,M_{\odot}$.  The evolution of galaxies in
these clusters provides information on the fate of the earliest objects
that formed in the Universe, and the observed mass-dependent evolution
of the abundance of clusters over $1<z<2$ provides strong constraints
on cosmological parameters (particularly $w$).
 
\begin{figure}[h]
\resizebox{0.65\textwidth}{!}{\includegraphics{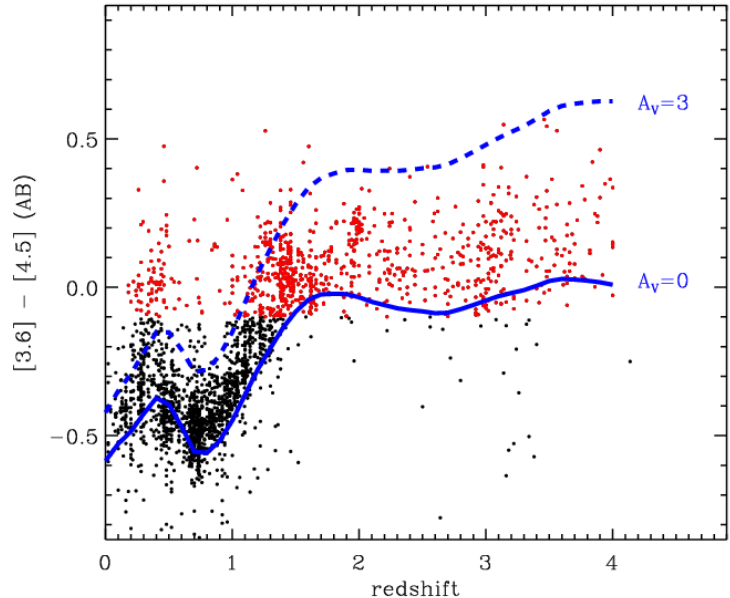}}
\caption{
The observed $[3.6]-[4.5]$ colors of galaxies versus redshift in the
GOODS-south field. The points show a sample of $K$-selected galaxies
(Wuyts \etal~in prep). The tracks show the
envelope of colors spanned by stellar population models with a
range of dust attenuations. A simple cut only in $[3.6]-[4.5]$ can
efficiently isolate galaxies at $z>1$ regardless of SED shape.
}
\label{iraccol.plot}
\end{figure}

IRAC 3.6 and 4.5 $\mu$m data alone can provide a crude redshift estimate,
as the $[3.6] - [4.5]$ color fairly cleanly separates galaxies with
redshifts below or above 1 (see Fig.\ \ref{iraccol.plot}).  However, the
returns from this survey will be greatly enhanced when it is performed in
an area, or areas, of sky with existing or planned acillary data. Examples
of such areas are the South Pole Telescope's SZ survey, and the fields
imaged by the near-IR VISTA Kilo Degree Survey (KIDS). The combination
of these data will not just allow detection of the clusters, but also
enable redshift- and mass estimates.
 
We note that wide-area, shallow surveys in high latitude fields
could also prove useful for Galactic programs, most notably for detecting
and characterizing the coldest brown dwarfs (e.g., \citet{Stern2007}).
Objects cooler than about 700 K, so-called ``Y dwarfs'', must exist.
Objects with inferred masses down to $\approx 5 M_{\rm Jup}$ have been
identified in star-forming regions and, according to theoretical models,
dwarfs less massive than $30 M_{\rm Jup}$ with ages $> 4$ Gyr should
have $T < 600$ K. However, none have been found to date.  This is
primarily because their SEDs peak at $\approx 4.5\,\mu$m (e.g., 
\citet{Burrows2003}), making them very faint at ground-based optical through
near-IR wavelengths.  For instance, a 600 K brown dwarf could only be
detected in the 2MASS PSC to about 1 pc.  In contrast, it
would be detectable in a 120 sec IRAC $\approx 4.5\,\mu$m image out to about
50 pc.

\section{Other Programs}
\label{othersurveys.sec}

We consider the three-tiered approach outlined above
an excellent starting point for designing observational programs
for studies of the distant Universe in the
warm era. Many other survey programs could, of course, be
considered, and we briefly discuss several alternatives here.

\subsection{A Medium Deep Survey Over Several 10s of deg$^2$}

There is a conspicuous gap in the three surveys discussed in this
document, as we jumped from a 20 hr depth over several deg$^2$
to a 120\,s depth over 100s of deg$^2$. A survey over
several 10s of deg$^2$ to a $\sim 1$\,hr depth would require
a similar investment as each of the surveys discussed in more
detail in the preceeding Sections. This territory is out of reach
of JWST and a unique niche for Spitzer in the warm era.

This committee failed to come up with a broadly defined, high impact
science case for this type of survey, but that may simply reflect
the biases and preconceptions of its members. A survey of this
type would map large scale structure at $z=2-4$ over a very
wide area, which could lead to new constraints on the growth
of dark matter halos and perhaps cosmology. Thanks to the
large number of galaxies that would be observed it would also
be possible to split the sample into many bins,
and study galaxy evolution as a function of luminosity, mass, color,
AGN-activity, and size.

\subsection{An Extremely Wide Survey of 1000s of deg$^2$}

It may seem odd to consider using a $5' \times 5'$ imager to cover areas
requiring hundreds of thousands of pointings. Nevertheless, the
unique wavelength regime and sensitivity of IRAC, combined with
the large amount of time that is potentially available in the
warm era, warrant a discussion of this question.

An ultra-wide survey will identify the most extreme objects
in the Universe, such as very luminous quasars and galaxy clusters
with masses $\sim 10^{15}\,M_{\odot}$.
However, the high overheads associated with very short integrations
make such a program either very inefficient or extremely
costly. The spacecraft overheads are such that they start
to dominate over the on-sky time for integration times per
pointing significantly shorter than $\sim 100$\,s.
As an example, a survey of 125 square degrees with a 120\,s
exposure time (comprising 4 dithered 30\,s exposures) takes
about 1000 hrs. A survey of 2500 square degrees with
a 6\,s exposure time (comprising 3 dithered 2\,s exposures)
would have the same total on-sky integration time, but cost
more than 6000 hrs due to greatly increased overheads.

Taking 120\,s exposure time as a minimum,
covering 2000 square degrees would be extremely costly as it would
require 16,000 hours. Such a large expenditure
may be difficult to
justify given the somewhat limited additional science that can
be accomplished above and beyond
the 500 deg$^2$ scenario discussed earlier.

\subsection{Gravitationally Lensed Galaxies}

Gravitational lensing by foreground clusters allows the study of
high redshift
galaxies fainter than the limits achievable in unmagnified fields
(e.g., \citet{Ellis2001, Stark2007b}),
and detailed analysis of
intrinsically more luminous galaxies (e.g., \citet{Pettini2000}).
The gain in S/N is substantial: the exposure time needed to
reach a given lensing-corrected limiting magnitude decreases
as $A^{-2}$ (for pointsources),
with the lensing amplification $A$ reaching
values of 20 in extreme cases.

This technique
has great potential, although there are some drawbacks:
the small volume that is sampled at high redshift (as
the relevant region is limited to a $\sim 1'$ diameter annulus whose
lensing-corrected area decreases with $A$), the requirement
that the mass distribution in the inner parts of the cluster can be
adequately modeled (to correct the measured properties for the
effects of lensing), and crowding. The latter aspect is particularly
problematic for IRAC, due to its large PSF compared to the distances
between galaxies in the central parts of clusters.

IRAC has already yielded interesting results in this area:
\citet{Egami2005} report the detection of a significant
Balmer break in a previously identified lensed
$z\sim 6.7$ galaxy, based on 3.6 and 4.5\,$\mu$m IRAC data of the
well-studied cluster Abell 2218. A program is currently underway
to systematically image 30 lensing clusters with IRAC, and
it may be very interesting to extend this type
of work in the warm era.

\subsection{The Stellar Populations of $z=10$ Galaxies}
\label{z10.sec}

Although these things are difficult to predict, it seems likely that
WFC3 on HST, VISTA, HAWKEYE on the VLT, or some
other new capability will identify a robust sample of $J$ band dropouts
in the near future
(see \citet{Bouwens2005}). 
IRAC imaging of these objects will
both confirm them (by establishing whether they have a blue
continuum redward of Ly$\alpha$) and constrain their stellar
populations by measuring the strength of the redshifted Balmer break
(which falls between the 3.6 and 4.5\,$\mu$m bands at this redshift).
We note that
extremely deep IRAC imaging may already be available if the objects
are found in a combined WFC3/IRAC survey,
as advocated above.

\subsection{Future Priorities}

The program described in the preceeding paragraph is
an example of science that cannot currently
be planned (although anticipated), and it is almost certain that
many other exciting possibilities will emerge during the remaining
lifetime of Spitzer. Such future programs can be large surveys,
but could also be small, very high impact observations of
special objects, special sky areas, or time-variable objects
(e.g., a $z=10$ gamma-ray burst).

It is crucial that a fraction of
the time available in the warm period
will remain unassigned, to accomodate
the shifting frontiers in the field. However, there will
be limitations
imposed by the anticipated reduction in user support.
It may be possible to have a TAC process twice during the
5 year warm mission
(rather than yearly) to assign remaining survey time
and to accomodate a limited number of
small, high-impact programs which do not require a large
support effort on the part of the SSC.

\section{Conclusions}

The end of Spitzer's cryogenic lifetime will leave its most sensitive
and versatile capability for studying the distant Universe intact,
enabling very ambitious survey programs addressing a wide
range of science.
Nearly anything that is done in the warm mission will explore unique
parameter space, as there is no competitive instrument in this
wavelength regime until JWST. Among the various possibilities,
we feel that the three-tiered approach outlined in this document
would extend currently
available samples by at least an order of magnitude, enable
qualitatively new science, and serve a wide community.
The survey parameters are summarized in Table 1.

\begin{center}
{ {\sc TABLE 1} \\
Recommended Surveys} \\
\vspace{0.1cm}
\begin{tabular}{cccll}
\hline
\hline
Area & Depth & Total time & Fields & Main science drivers \\
\hline
150\,arcm$^2$ & 250\,hr & $\sim 2500$\,hr & TBD & galaxies at
$z=7-14$\\
2\,deg$^2$ &  20\,hr & $\sim 7500$\,hr & COSMOS, UDS, & bright galaxies
at $z>6$\\
                &          &                &EGS, E-CDFS   & AGN at $z=1-7+$\\
                &          &                & & clustering at $z=2-6$\\
500\,deg$^2$ & 120\,s & $\sim 4000$\,hr & TBD & quasars to $z\sim 7$ \\
              &               &          &    & galaxy clusters at $1<z<2$\\
\hline
\end{tabular}
\end{center}

Chosing survey fields is a charged subject, as several large groups
in the high redshift community have invested significant
effort and resources in particular areas of the sky. This
document leaves
this issue open for the ultra-deep
and shallow  surveys, as there
are no fields that can be easily identified as superior to
all others. Distributing these surveys may also be an option, e.g.,
covering two widely separated 75\,arcmin$^2$ fields in the ultra-deep
survey rather than a single 150\,arcmin$^2$ area.

However, we are explicit about the fields that can be covered
in the deep 2 deg$^2$ survey. Despite the large investment
of IRAC time that would be required, we
believe a case can be made for covering all four well-studied
$>0.25$ deg$^2$ fields. Each of these fields offers qualitatively
different legacy value: the UDS and COSMOS fields will have
the best near-IR coverage, the EGS has the best spectroscopy,
a lower mid-IR background than the equatorial fields, and
is well placed for Northern telescopes,
and the E-CDFS has very low mid-IR background
and is ideally placed for Chilean telescopes (including ALMA).
These four fields have been vetted for their legacy
value by many
time allocation committees for ground- and space-based
facilities, and one may question
whether it is sensible to do that yet again.

An important consideration in this context is not just the
quality of the supporting data, but their access.
The survey programs that are
considered in this document require such a large investment of Spitzer
time that a level playing field is absolutely crucial. A field
should therefore only be observed if access
to crucial supporting data (e.g., near-IR imaging) is completely
unrestricted. This would be an extension of the usual process,
where proposers use their (often partially proprietary) data to
argue for a certain survey strategy or sky area, and then promise
to make the space-based data publicly available in reduced form.

The TAC process is also unusual, in the sense that the size of the
envisioned proposals will exceed even the largest programs that have
been executed on space observatories to date. It is
unlikely that proposers will have a chance to revise their proposals
for a future round, as a large fraction of the available time over
the entire warm mission may be reserved in a single proposal round.
TACs inevitably vary in their composition, priorities, and expertise,
and special care needs to be taken to ensure that the best science
is selected for this unique opportunity.

In practice it may be desirable to have the
SSC implement one or more of the TAC-approved
surveys, following the example
of the various Hubble deep field campaigns. This will
capitalize on the experience and expertise of the SSC staff,
ensure a timely distribution of reduced data, and allow the
community to focus their efforts on the science enabled by these
surveys rather than their execution.

\begin{theacknowledgments}
We thank
Rychard Bouwens,  Ranga-Ram Chary, 
Alastair Edge, Eiichi Egami,
Jonathan Gardner, Subha Majumdar,
and Adam Stanford for their input, and Lisa Storrie-Lombardi for
her expert coordination of the Warm Mission preparatory
efforts.
\end{theacknowledgments}

\end{document}